\pdfoutput=1

\documentclass[11pt]{article}

\usepackage[]{ACL2023}

\usepackage{times}
\usepackage{latexsym}
\usepackage{graphicx}
\usepackage{amsmath}
\usepackage{amssymb}
\usepackage{multirow}
\usepackage{amsfonts}
\usepackage{fdsymbol}
\usepackage{enumitem}
\setlist{leftmargin=*}
\usepackage{flafter}
\usepackage{tabularx}
\usepackage{listings, xcolor}
\usepackage[T1]{fontenc}

\usepackage[utf8]{inputenc}

\usepackage{microtype}

\usepackage{inconsolata}
\usepackage{booktabs}
\usepackage{pifont}
\newcommand{\cmark}{\ding{51}}%
\newcommand{\methodnamews}{\text{HierarchyNet}}
\newcommand{\methodname}{\methodnamews~}

\title{HierarchyNet: Learning to Summarize Source Code  with \\ Heterogeneous Representations}

\author{
	Minh Huynh Nguyen$^{\clubsuit}$\thanks{\ \ Equal contribution. Listing order is based on the alphabetical ordering of author surnames.}, \  Nghi D. Q. Bui$^{\vardiamondsuit}$\footnotemark[1], \ Hy Truong Son$^\spadesuit$, \ Long Tran Thanh$^{\varheartsuit}$, \ \textbf{Tien N. Nguyen$^{\diamondsuit}$}\\
	$^\clubsuit$ FPT Software AI Center,
	$^\vardiamondsuit$Department of Computer Science, Fulbright University, Viet Nam\\
	$^\spadesuit$Department of Computer Science, University of Chicago, IL, USA\\
	$^\varheartsuit$Department of Computer Science, University of Warwick, UK\\
	$^\diamondsuit$School of Engineering and Computer Science , The University of Texas at Dallas, USA\\
	\thanks{Emails: minhnh46@fpt.com.vn, dqnbui.2016@smu.edu.sg, hytruongson@uchicago.edu, long.tran-thanh@warwick.ac.uk, tien.n.nguyen@utdallas.edu}
}
\begin{document}
	\maketitle
	\vspace{0.5cm} 
	\begin{abstract}

We propose a novel method for code summarization utilizing Heterogeneous Code Representations (HCRs) and our specially designed HierarchyNet. HCRs effectively capture essential code features at lexical, syntactic, and semantic levels by abstracting coarse-grained code elements and incorporating fine-grained program elements in a hierarchical structure. Our HierarchyNet method processes each layer of the HCR separately through a unique combination of the Heterogeneous Graph Transformer, a Tree-based CNN, and a Transformer Encoder. This approach preserves dependencies between code elements and captures relations through a novel Hierarchical-Aware Cross Attention layer. Our method surpasses current state-of-the-art techniques, such as PA-Former, CAST, and NeuralCodeSum. 

	\end{abstract}
	\section{Introduction}
	\label{sec:introduction}

Code summarization is an important task for software developers as it helps them understand and maintain source code effectively. However, manual documentation of code can be a time-consuming and labor-intensive task for developers. To overcome this challenge, there is a need for an automated approach to generate code comments. To generate accurate summaries, a model must understand essential features in source code, including lexical, syntax, and semantics information. A crucial part
of semantic information is the relations among program elements including their data and control dependencies. Any code representation learning approach needs to represent well such information in order to perform well in the downstream tasks including code summarization.

Early sequence-based techniques~\cite{iyer2016codenn, ahmad2020ncs} treated code as a sequence of texts, but they did not take into account the complex interdependence of program elements in syntax or semantics. Structured-based~approa\-ches \cite{alon2018codeseq, leclair2019funcom,Shi2021CASTEC,chai2022pyramid} were later proposed to better capture the syntactic information. The state-of-the-art approaches, such as CAST~\cite{Shi2021CASTEC} and PA-Former~\cite{chai2022pyramid}, leverage the idea of \textit{hierarchically} splitting the AST into smaller parts based on its structure. CAST hierarchically splits the AST's code blocks based on certain attributes, while PA-Former works by treating statements as spans and splitting them into tokens and sub-tokens. These code-hierarchy approaches bring the benefits in terms of effective and affordable training of neural models. However, a common drawback is that they ignore the program dependencies in their representations. There are other lines of work based on graphs~\cite{leclair2020improved,fernandes2018structured,Hellendoorn2020great}  that models program dependencies by adding edges to the AST, in which the edges are the dependencies derived from static analysis. However, these approaches do not take into account the code hierarchy as the previous line of work. 

We propose a novel \textit{Heterogeneous Code Representation} (HCR) that addresses these limitations by \textit{combining the best of both worlds}. The key advantage of HCR is that it captures the essential code attributes at lexical, syntactic, and semantic levels in a code hierarchy that 
represents program elements based on their characteristics: sequences for code tokens, AST subtrees for syntax, and graphs for dependencies. Importantly, we effectively capture program dependencies by selectively abstracting coarse-grained nodes into a higher-level layer and fine-grained nodes into a lower-level layer. This contributes to better generating summaries as our model will have a better and more comprehensive understanding of the source code. 
To process our heterogeneous representations, we introduce a heterogeneous architecture, called \textit{\methodnamews}, which comprises a Transformer Encoder for processing lexical information, a Tree-based Encoder for processing syntactic information, and a Graph-based Encoder for capturing program dependencies. These layers do not process information individually but hierarchically, which intuitively captures the relationships between program elements even better. Additionally, we introduce the Hierarchy-Aware Cross Attention layer and a Gating layer to combine information across layers effectively.

We systematically conducts an evaluation on our proposed model, utilizing adaptations of HCR and \methodnamews, on various established datasets for code summarization, including TL-CodeSum~\cite{hu2018deepcode}, DeepCom \cite{hu2018deep}, and FunCom~\cite{leclair2019funcom}. The results demonstrate that our model surpasses the state-of-the-art baselines by a significant margin. 

To summarize, our key contributions include: 

\textbf{(1)} \textit{Heterogeneous Code Representation}: a novel code representation that models source code with sequences, trees, and graphs that fit with its characteristics at lexical, syntactic, and semantic levels. 

\textbf{(2)} {\textit{\methodnamews}}: 
a novel heterogeneous neural network architecture, designed in a modular manner, where each module in the architecture is responsible for processing each layer in the \textit{Heterogeneous Code Representation}. The key modules include the Transformer Encoder, Tree-based CNN, and Heterogeneous Graph Transformer, as well as a novel Hierarchy-Aware Cross Attention module for attending to information across layers.


\textbf{(3)} Our experiments show superior performance in comparison to baselines. Our model outperforms PA-Former, the state-of-the-art for DeepCom, by 4.79 BLEU points and outperforms CAST, the state-of-the-art for TL-CodeSum, by 2.82 BLEU points. Notably, these baselines only show a slight improvement of 1.36 BLEU points on average, while our model achieves a significant improvement of 5.175 BLEU points on the same baselines.

\section{Motivation}
\label{sec:motiv}

\begin{figure*}
	\centering
	\includegraphics[width=0.85\linewidth]{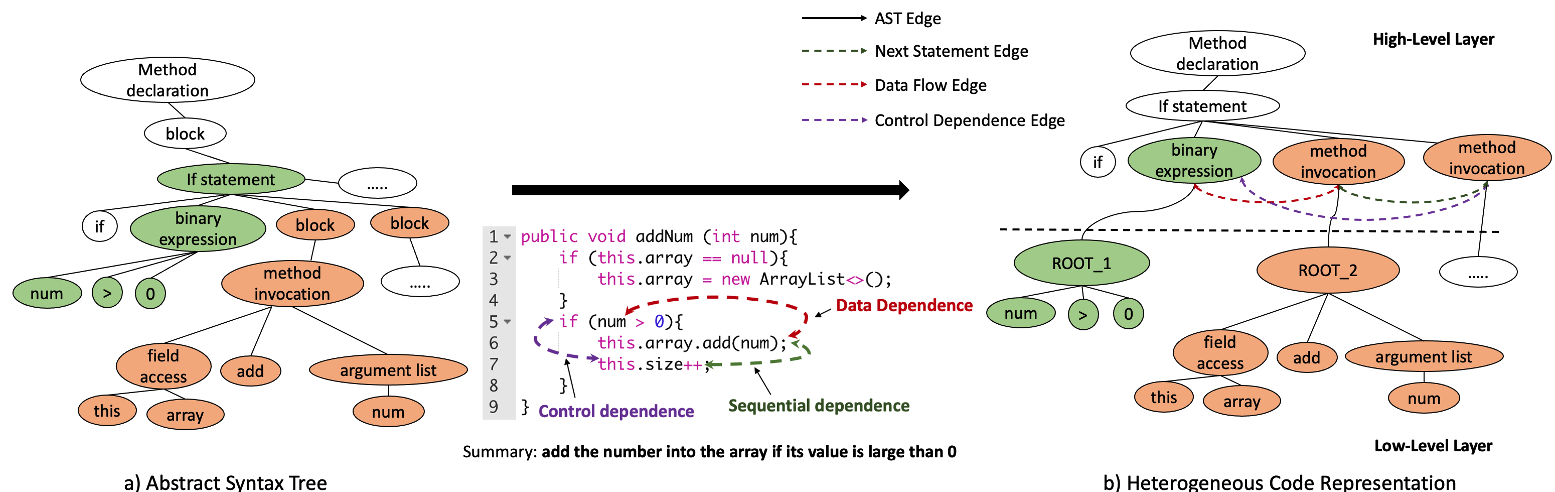}
	\caption{\small Motivation on the need of heterogenous code representation}
	\label{fig:motivation}
\end{figure*}

Let us use an example to motivate the key characteristics in our solution. Figure~\ref{fig:motivation} shows a code snippet and its corresponding summarization. The task is to collect the positive numbers into an array. To be effective in the summarization, a model needs to capture several code features at the lexical, syntactic, and semantic levels. For example, at the lexical level, the sub-tokens \texttt{add}, \texttt{num}, and \texttt{array} appear as a whole or closely resemble to the words in the summary. The lexical tokens \texttt{`$>$'} and \texttt{`0'} correspond to the texts \texttt{`larger than'} and \texttt{`0'} in the summary. At the syntactic level, a model needs to recognize the syntactic structures of the code. For example, the \texttt{if} statement at line 5 could help a model put a higher score to a conditional sentence in the summary texts. 

Importantly, the control and data dependencies among the statements could provide a good indication on the intention of the execution orders of the statements. Without considering the control dependencies between the statements, a model cannot capture such intention because the sequential order in the code might not reflect the execution order among the statements. For example, despite their sequential order, the execution of the statement at line 6 is not guaranteed to follow the statement at line 5, as it is dependent on the outcome of the \textit{if condition} at line 5. 
Moreover, the data dependency via the variable \texttt{num} at line 5 and line 6 is also important for code summarization because both appearances referring to the same variable could allow a model to understand that only the numbers satisfying the positive condition are collected. 



Previous approaches, such as those outlined in \citet{leclair2020improved}, \citet{fernandes2018structured}, and \citet{Hellendoorn2020great}, have utilized heuristics from static analysis to connect nodes in the AST to represent dependencies. However, the large size of the AST can make it difficult for a model to effectively capture dependencies among nodes that are far apart~\cite{alon2021oversquashing}. In contrast, state-of-the-art approaches such as CAST~\cite{Shi2021CASTEC} and PA-Former~\cite{chai2022pyramid} aim to create a hierarchy among code elements by splitting the AST into smaller parts, flattening each part into a sequence, and feeding it into sequence-based models for processing. However, this method does not preserve the structural and semantic relations among program elements and does not maintain program dependencies among the elements.




We propose the \textit{Heterogeneous Code Representation} (as depicted in Figure~\ref{fig:motivation}b) is a way to restructure code into different layers, abstracting meaningful entities such as statements or expressions into single nodes in a higher layer. As seen, the dependencies between elements can be easily represented in HCR, with the data dependence edge (red dashed line) connecting the coarse-grained nodes \textit{binary\_expression} (line 5) and \textit{method\_invocation} (line 6), the control dependence edge (purple dashed line) connecting the coarse-grained nodes \textit{binary\_expression} (line 5) and \textit{method\_invocation} (line 7), and the next-statement edge (sequential dependence) connecting \textit{method\_invocation} nodes (statements at line 6 and 7). Also, the AST edge that captures program's syntax is also preserved at both the low-level and high-level layers.

Our code hierarchy with a heterogeneous representation is suitable to represent code features at all three abstraction levels with sequence, tree-structured, and graph-structured representations. We also have a heterogeneous neural network framework to handle our heterogeneous representation. At each level, we use an appropriate neural network: a sequence-based transformer for lexical code tokens,
a tree-based encoder for AST subtrees, and a heterogeneous graph transformer for the coarse-grained dependencies. In contrast, CAST~\cite{Shi2021CASTEC} and PA-Former~\cite{chai2022pyramid},
have an un-natural code hierarchies to fit with their monolithic neural networks (see Appendix). 
Moreover, our approach reduces the number of nodes to process at the high level (Figure~\ref{fig:motivation}), thus reducing the computational workload in the corresponding neural network, as the computation of a larger number of finer-grained nodes is offloaded to another neural network. It also better captures the dependencies between far-away nodes in an AST, which is an issue with the prior works. 

\section{Related Work}
\paragraph{Code Summarization} Generating accurate descriptions for source code has been a topic of ongoing research. In the early stages, sequence-based techniques were utilized to process code similar to text~\cite{iyer2016codenn, ahmad2020ncs, wei2019code}, without considering the dependencies of program elements through syntax or semantics. For example, NeuralCodeSum~\cite{ahmad2020ncs} is a purely transformer-based approach that receives the sequence of code tokens and generates summaries. Later, structure-based and tree-based approaches were proposed to capture the syntax of the source code~\cite{TreeLSTM, TBCNN, bui2021treecaps, leclair2019funcom, hu2020deep, peng2021tptrans, Shi2021CASTEC, chai2022pyramid}. For instance, TreeLSTM~\cite{TreeLSTM} works by accumulating node information in a bottom-up style, while TPTrans~\cite{peng2021tptrans} integrates path information from the AST into the transformer. CAST~\cite{Shi2021CASTEC} and PA-former~\cite{chai2022pyramid} are currently the state-of-the-art methods with the same key idea of breaking the code into a structural hierarchy. Finally, graph-based techniques were used to capture code semantics by adding inductive bias into the AST through semantic edges, turning it into a graph~\cite{leclair2020improved,fernandes2018structured,Hellendoorn2020great}. However, they still suffer issues in the representations of code hierarchy and program dependencies, as well as neural networks to handle them (see Appendix for details). 


\paragraph{Pretrained Language Models of Code}
Apart from code summarization, language models of code generally support a wide range of code understanding tasks, such as code generation~\cite{codebert,codet5,elnaggar2021codetrans}, code completion~\cite{codebert,codet5,peng2021could}, program repair~\cite{xia2022practical}, code translation~\cite{roziere2020unsupervised} and so on. A large body of recent work employs language models from natural language processing for code~\cite{codebert,codet5,  guo2020graphcodebert, ahmad2021unified, bui2021self, elnaggar2021codetrans,peng2021could, kanade2020learning, chakraborty2022natgen, ahmed2022multilingual, niu2022spt}. They mostly treat code similar to texts and adapt the same pretraining strategies as for natural languages. CodeBERT~\cite{codebert} adapts a Roberta model~\cite{liu2019roberta} to pretrain a model of code on multiple programming languages. GraphCodeBERT~\cite{guo2020graphcodebert} uses a data-flow pretraining strategy. CodeT5~\cite{codet5} extracts unique identifier information from source code to pretrain the T5~\cite{raffel2019exploring} model for code in a multi-modal style.

\section{Overview}
\label{sec:overview}

\begin{figure*}
	\centering
	\includegraphics[width=1.0\linewidth]{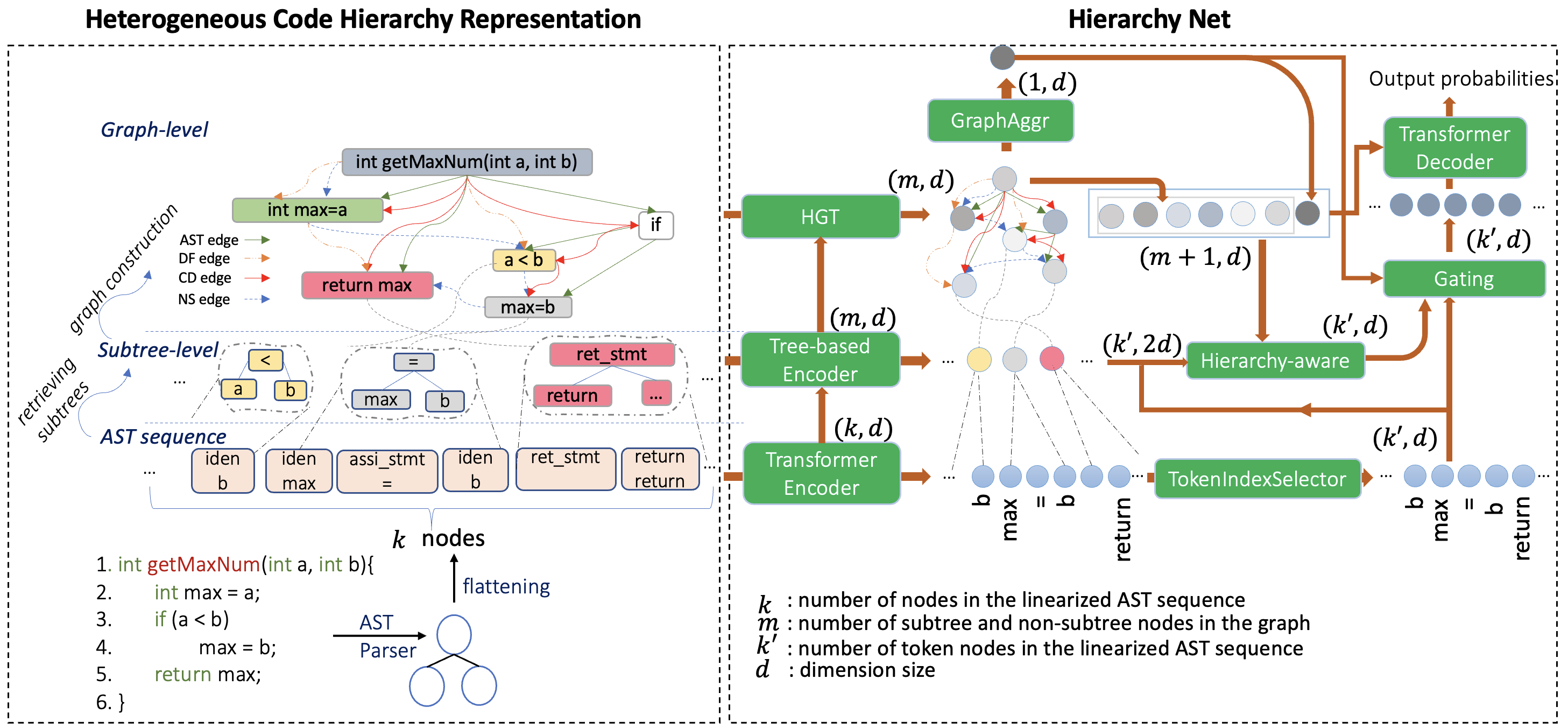}
 \vspace{-18pt}
	\caption{\small {\methodnamews} Architecture}
	\label{fig:overview}
\end{figure*}

Figure~\ref{fig:overview} shows an overview of our model with the following key ideas: 
\begin{enumerate}[leftmargin=*]
	\item Heterogeneous Code Representations (HCRs) for multi-level code features and dependencies.
	\item {\methodname} to process each layer in the HCRs separately and effectively.
\end{enumerate}

The HCRs and the {\methodname} architecture are designed to work in tandem, with each layer of {\methodname} processing the corresponding layer of the HCRs. The HCRs provide an effective representation of program dependencies by using separate representations for sequences, trees, and graphs at different levels: lexical, syntactic, and semantic. We focus on two key elements of a program: statements and expressions.
At the lowest level, the AST-sequence layer explicitly represents the sequences of code tokens and is handled by the Transformer Decoder~\cite{NIPS2017_attention}. This layer processes node-token and node-type embeddings to initialize the nodes in the next level, the Subtree-level layer. The subtree-level layer captures the syntactic structures among statements and expressions and is represented as AST subtrees. Each subtree represents an abstract and meaningful node in the higher level, the Graph-level layer. The graph-level layer contains a variety of edge types, including data flow and control dependence, and is used to describe the dependencies and relations among statements and expressions. This layer is processed by the Heterogeneous Graph Transformer (HGT) \cite{hgt}. In addition, we also propose two novel layers to enhance expressive power of \methodnamews, which are Hierarchy-aware Cross Attention and the Gating Mechanism.

\section{Heterogeneous Code Representation} \label{sec:code-hierarchy-rep}

We begin by parsing a program into an Abstract Syntax Tree (AST) $T$. This AST is then serialized to create a sequence of nodes $L$. On the other hand, a set of subtrees at the statement and expression level are extracted, resulting in a new, significantly smaller AST $T'$. A graph $G$ for program dependencies is built by utilizing these subtrees and applying static program analysis. This final representation, referred to as the "Heterogeneous Code Representations" is illustrated in Figure \ref{fig:overview} (left-side) and can be divided into three layers. The first, called the "Linearized AST Sequence," is a sequence of nodes from the serialized AST. The second layer, the "Subtree-level," represents statement and expression-level program elements, each represented by a subtree consisting of nodes from the original AST $T$. The highest-level and coarsest-grained layer, the "Graph level," is a graph $G$ consisting of nodes from $T'$, enriched by semantic edges such as control and data dependencies. Next, we present in details each layer in our model.


\subsection{Serialized AST Sequence}
%



Each token node contains a non-empty token, which is often made up of multiple sub-tokens. To incorporate these sub-tokens, we insert new sub-token nodes as children of the corresponding token node. We convert the AST into a sequence of nodes by traversing such that the original token order is maintained (Figure \ref{fig:overview}). Formally, the linearized AST sequence $L = [l_1, l_2, ..., l_{k}]$ (where $k$ is the size of $T$) represents the lowest level of HCRs.



\subsection{Syntactic Level}

A function is usually a combination of many statements and expressions, each of which often represents a sufficient amount of information to understand how/what it does. We extract the AST subtrees corresponding to statements and expressions.
The set of subtree types can be looked up in the Appendix.
These subtrees are then abstracted by replacing them with placeholder nodes in $T$, resulting in a smaller tree $T^\prime$. 
This process is done through a depth-first traversal of the AST, where subtrees are replaced and further traversal is halted at the subtree's root node. This results in a new tree $T^\prime$ and a set of subtrees $ST$, with some nodes in $T^\prime$ pointing to elements in $ST$, which forms the second level in our HCRs. Note: some nodes in $L$ do not belong to any subtrees (non-subtree nodes).
\subsection{Semantic Level}
\label{subsec:semantic-level}


 
 We use the reduced AST $T'$ and incorporate semantic edges among the nodes to create graph $G$ (as depicted in Figure \ref{fig:overview}). Our graph includes four distinct edge types: AST edges, Data-flow (DF) edges, Control-dependence (CD) edges, and Next-subtree (NS) edges. These edges represent various forms of connections between program elements, such as code structures, data and control dependencies, and sibling statements. Further explanations of each semantic edge can be found in the Appendix.

\section{Neural Network Architecture} \label{sec:model-architecture}


This section explains the \methodname method for processing code hierarchies (Figure \ref{fig:overview}). Each node $l_i \in L$ has two attributes: \textit{token} and \textit{type}. The initial representation of each node $l_i$ is computed by concatenating the embeddings of its \textit{token} and its \textit{type}. These embeddings can be looked up from two learnable embedding matrices (token and type). We denote $s_i$ be the initial embedding of the node $l_i$, $i \in \mathbb{N}, 0 < i \le k$ where $k$ is the length of $L$.

\subsection{Transformer Encoder}

The Transformer Encoder encodes the linearized AST sequence $L$ to capture lexical values. It takes initial embeddings $[s_1, s_2, ..., s_{k}]$ as input and produces the output $[h_1, h_2, ..., h_{k}]$ .

\subsection{Tree-based Encoder}
This layer's primary role is to process the subtrees in the Subtree layer. Additionally, it also embeds non-subtree nodes in the $L$ by applying a non-linear transformation. To model local patterns and hierarchical relations among nodes within the same subtree, all subtrees are passed through a Tree-based CNN \cite{mou2016convolutional}. An attention aggregation method~\cite{Alon2019} is then employed to encode each subtree as an embedding vector, using a global attention vector $\alpha$. The output of this layer are denoted as $\{ \hat{t}_i\}_{i=1}^{m}$ where $m$ is the number of subtrees and non-subtree nodes.

\subsection{Heterogeneous Graph Transformer (HGT)}
After obtaining the embeddings of all the subtrees, we further encode the dependencies among the nodes in the heterogeneous graph $G$. We adapt the Heterogeneous Graph Transformer (HGT) \cite{hgt} to process the graph effectively. The outputs are the vectors $\{n_i\}_{i=1}^{m}$ that not only bring textual information (by Transformer Encoder and next-subtree edges) but also are contextualized by the locally hierarchical structures of the subtrees and dependence information that are unique characteristics in source code.
\subsection{Graph Aggregation (GraphAggr)}
Upon completion of the HGT processing, it is necessary to aggregate the individual nodes within the graph into a vector that represents for the graph. Similar to the tree aggregation techniques employed in the Tree-based Encoder, an attention mechanism is utilized to aggregate the nodes and generate a graph embedding, represented as $g$. This graph embedding encapsulates the overall semantic meaning of the code.
\begin{equation*}
g = \sum_{i=1}^{m}\frac{\exp{\beta^T{n}_i}}{\sum_{j=1}^{m}\exp{\beta^T{n}_j}}\cdot{n}_i
\end{equation*}
where  $\beta$ is the global attention vector, and $\{n_i\}_{i=1}^{m}$ are the output of HGT.
\subsection{Token Index Selector}

The TokenIndexSelector layer utilizes the output of the Transformer Encoder as input and serves to retain the embeddings of nodes $l_{i}$ that possess non-empty token attributes while discarding those that do not. The rationale is that the Transformer Encoder effectively encodes textual meaning but is inadequate in encoding syntax (as represented by the type attribute), which could potentially introduce noise to subsequent layers (such as the Gating Layer and Transformer Decoder). It is worth noting that the Subtree layer effectively encodes syntax information using Tree-based CNN. Formally, let $H'$ be the sequence of the elements $h_i$ such that $l_i$ is a token node, for all $0 < i \le k$. We denote the members of $H'$ by $h'_1, h'_2, \dots, h'_{k'}$ where $k'$ is the number of token nodes in the $L$.

\subsection{Hierarchy-Aware Cross Attention (HACA)}


Although information is gathered in a bottom-up manner, there may still be missing connections between layers. To address this issue, we introduce the Hierarchy-aware Cross Attention layer, which calculates the attention of each token towards nodes in the structure (tree + graph). This layer, illustrated in Figure~\ref{fig:overview}, enables the TokenIndexSelector layer to focus on information in the HGT layer. The keys $K$ and values $V$ in this layer are derived from the combination of the nodes' embeddings $\{n_i\}_{i=1}^{m}$ and the graph embedding $g$. Additionally, a token can occur multiple times in a code snippet and, even with the use of positional encoding, the vectors of these tokens may be similar. To address this issue, we concatenate the subtrees of these occurrences in order to distinguish them. For example, by examining the corresponding subtrees, we can discern the different roles of the variable $a$ at lines 1 and 2. This allows us to enhance  the discrepancies of the same token in different positions, we concatenate $h'_i$ and $\hat{t}_i$ to create the vector query $q_i$; formally, $q_i = f_{ca}([h'_i, \hat{t}_i])$ where $f_{ca}$ is a projection from $\mathbb{R}^{2d}$ to $\mathbb{R}^d$ . Then the cross-attention is computed as usual, that is $\text{softmax} \left( \frac{QK^T}{\sqrt{d_k}} \right)V$ where $d_k$ is the inner dimension size of each attention layer. This layer produces $\{c_i\}_{i=1}^{k'}$ where $c_i$ is the fused hierarchical context dedicated to the token node corresponding to $h'_i$, for all $ 0 < i \le k'$.

%

\subsection{Gating Layer}

The HACA layer is responsible for calculating attention scores across different layers, but it does not perform any combination of the information.  We introduce the Gating layer to combine the information across different layers in the hierarchy, which serves as the input for the Transformer Decoder. The goal is to combine the outputs $\{c_i\}_{i=1}^{k'}$ of HACA with the lexical information of $\{h'_i\}_{i=1}^{k'}$. To balance the two sources of information, we propose to add a sufficient amount of context from ${c_i}$ to ${h'_i}$. We take inspiration from the gating layer in \citet{cho-etal-2014-properties}, and modify it to achieve this goal. Specifically, the ratio between the two sources is controlled by the graph embedding, as $g$ is the highest level of abstraction and contains a global understanding of the code. Formally, the computation can be summarized as: $\lambda = \text{sigmoid}(Wg + b)$\label{eq:gating}, where $W \in \mathbb{R}^{d \times d}$, $b \in \mathbb{R}^d$ or  $W \in \mathbb{R}^{d}$, $b \in \mathbb{R}$ are learnable parameters, and $d$ is the dimension of the vector $g$. 
We then apply a non-linear projection $f_c$ to map $c_i$ onto the space of $h'_i$ and form the hierarchy-aware hidden state by: 
\begin{equation*}
e_i = \lambda f_c(c_i) + (1 - \lambda)h'_i
\end{equation*}

Finally, $\{e_i\}_{i=1}^{k'}$ are final encoder hidden states.

\begin{table*}[t]
	\centering
	\fontsize{8.5}{8.5}\selectfont 
	\begin{tabular}{ c c c c c c c c c c c c } 
		\toprule
		\multirow{2}{*}{Model} & \multicolumn{4}{c}{TL-CodeSum}  & \multicolumn{4}{c}{FunCom} \\
		\cmidrule(lr){2-5} \cmidrule(lr){6-9}
		& BLEU & Meteor & Rouge-L & Cider & BLEU & Meteor & Rouge-L & Cider\\
		\midrule
		Code2seq & 16.09 & 8.94 & 24.21 & 0.66  & 23.84 & 13.84 & 33.65  &1.31\\
		CodeNN & 22.22 & 14.08 & 33.14 & 1.67 & 20.93 & 11.44 & 29.09  &0.90\\
		HDeepCom & 23.32 & 13.76 & 33.94 & 1.74 & 25.71 & 15.59 & 36.07  & 1.42\\
		HybridDrl & 23.51 & 15.38 & 33.86 & 1.55  & 23.23 & 12.25 & 32.04  & 1.11\\
		AttGru & 29.72 & 17.03 & 38.49 & 2.35  & 27.82 & 18.10 & 39.20 & 1.84  \\
		ASTAttGru & 30.78 & 17.35 & 39.94 & 2.31 & 28.17 & 18.43 & 39.56 &1.90\\
		NCS & 40.63 & 24.86 & 52.00 & 3.47  & 29.18 & 19.94 & 40.09  & 2.15\\
		CodeAstnn & 41.08 & 24.95 & 51.67 &3.49 & 28.27 & 18.86 & 40.34 &1.94\\
		\midrule
		CAST & 45.19 & 27.88 & 55.08 & 3.95 & 30.83 & 20.96& 42.71  & 2.31\\
		\midrule
		\methodname & \textbf{48.01} & \textbf{30.30} & \textbf{57.90} &  \textbf{4.20} &\textbf{32.52} & \textbf{21.48} & \textbf{42.76} & \textbf{2.33} \\
		\bottomrule
	\end{tabular}
	\caption{\small Results on code summarization on TL-CodeSum and FunCom. Our method outperforms CAST and other baselines with significant margins in terms of BLEU, Meteor, Rough-L and Cider.}
	\label{tab:code-summarization-cast}
\end{table*}
\begin{table*}[t]
	\centering
	\fontsize{8.5}{8.5}\selectfont 
	\begin{tabular}{c c c c c c c c c  } 
		\toprule
		\multirow{2}{*}{Model} & \multicolumn{4}{c}{DeepCom}  & \multicolumn{4}{c}{FunCom-50} \\
		\cmidrule(lr){2-5} \cmidrule(lr){6-9}
		& BLEU & Meteor & Rouge-L & F1  & BLEU & Meteor& Rouge-L & F1  \\
		\midrule
		CodeNN & 28.45 & 17.89 & 43.51  & 44.77 & 31.86 & 19.11 & 48.90 & 49.92  \\
		TreeLSTM & 28.99 & 18.18 & 43.99 & 45.29 & 31.46 & 18.87 & 48.29 & 49.33\\
		HDeepCom & 32.18 & 21.53 & 49.03 & 50.75  & 35.06 & 22.65 & 53.35  & 54.81 \\
		ASTAttGru & 33.04 & 22.21 & 49.76 & 51.47  & 37.00 & 23.75 & 55.03 & 56.52 \\
		
		SiT & 35.69 & 24.20 & 53.75 & 55.72 & 42.12 & 26.82 & 59.33 & 60.84\\
		GREAT & 36.38 & 24.18 & 53.61 & 55.46 & 43.29 & 27.44 & 60.36  & 61.83\\
		NCS & 37.13 & 25.05 & 54.80  & 56.68 & 43.36 & 27.54 & 60.41  & 61.86\\
		TPTrans& 37.25 & 25.02 & 55.00 & 56.88 & 43.45 & 27.61& 60.57 & 62.03\\
		CAST & 37.20 & 25.07 & 54.87 &56.75  & 43.58 & 27.67& 60.52  & 61.98\\
		\midrule
		PA-former & 38.85 & 25.90 & 56.10 &  57.90 & 44.65 & 28.27 & 61.45  & 62.86 \\
		\midrule
		\methodname & \textbf{43.64} & \textbf{29.22} & \textbf{59.00} & \textbf{60.53}&  \textbf{51.12}  &  \textbf{34.13}  &  \textbf{65.43}    &  \textbf{66.64}    \\
		\bottomrule
	\end{tabular}
	\caption{\small Results on code summarization on DeepCom and FunCom-50. Our method outperforms PA-former and other baselines with significant margins in terms of BLEU, Meteor, Rough-L and F1.}
	\label{tab:code-summarization-pa}
\end{table*}

\subsection{Transformer Decoder}

Unlike the vanilla Transformer decoder \cite{NIPS2017_attention}, we need to combine the two sources, including hierarchy-aware textual information (the output of Gating layer) and the structural/semantic meaning (the output of both HGT and GraphAggr). The serial strategy \cite{serial-decoder} is adopted, which computes the encoder-decoder attention one by one for each input encoder. The key and value sets of the first cross-attention come from the output of HGT and GraphAggr. Those sets of the other cross-attention are from the output of Gating layer.

\section{Empirical Evaluation}
\label{sec:eval}
\begin{table*}[t]
	
	\resizebox{\textwidth}{!}{
		\begin{tabular}{@{}lcccccccccc@{}}
			\toprule
			\multirow{2}{*}{ID} & 	\multirow{2}{*}{Tokens} & \multirow{2}{*}{Subtrees} & \multicolumn{4}{c}{Graph}                  & \multirow{2}{*}{BLEU} & \multirow{2}{*}{Meteor} & \multirow{2}{*}{Rouge-L} & \multirow{2}{*}{Cider} \\ \cmidrule(lr){4-7}
			&               &            & AST edges & NS edges & CD edges & DF edges &                       &                         &                          &                        \\ \midrule
		1 &	\cmark	&         -                 &      -       &      -      &     -       &      -      &             40.63          &      24.86                   &        52.00                  &               3.47         \\
		2&	\cmark	&    \cmark                  &      -       &      -      &     -       &      -      &             44.16         &      28.19                  &        55.48                  &               3.77 \\  
		3&	\cmark	&         \cmark                  &     \cmark        &      -      &     -       &      -      &             45.37         &      28.43                 &        55.72                  &               3.91\\  
			4&	\cmark	&         \cmark                  &      \cmark        &      -      &   \cmark        &      -      &             46.54         &      29.39                   &        56.70                  &               4.04\\  
				5&	\cmark	&         \cmark                  &      \cmark        &      -      &     -       &     \cmark       &             46.61        &      29.41                  &        56.64                  &               4.03\\  
					6&	\cmark	&         \cmark                  &      \cmark        &      -      &    \cmark        &      \cmark       &             47.46        &      30.15                  &        57.63               &               4.14\\  
			7&\cmark	&         \cmark                  &      \cmark        &    \cmark       &   -       &     -      &             45.44        &      28.24                  &        54.72             &               3.89\\  
			8&\cmark	&         \cmark                  &      \cmark        &      \cmark       &     \cmark        &      -      &             46.84        &      29.40                  &        56.88             &               4.05\\  
			9&\cmark	&         \cmark                  &      \cmark        &      \cmark       &   -       &      \cmark       &             47.26        &      30.10                 &        57.64           &               4.12\\  
			\midrule
		10 &	\cmark	&         \cmark                  &      \cmark        &      \cmark       &     \cmark        &      \cmark       &             \textbf{48.01}       &        \textbf{30.30}                &          \textbf{57.90}            &                 \textbf{4.20} \\  
			\bottomrule
	\end{tabular}}
	\caption{\small Results of ablation study}
	\centering
	\label{tab:ablation-bottomup}
\end{table*}

\textbf{Experiment Settings.} We implemented our model based using Pytorch.  We selected AdamW \cite{loshchilov2018adamw} as the optimizer, and we used a linear warm-up learning rate scheduler to stabilize the training process. We also adopted early stopping with patience 20. Other hyper-parameters are described in the Appendix.



\textbf{Datasets.} We aimed to conduct a comprehensive comparison of the state-of-the-art (SOTA) baselines. Note that different baselines use distinct datasets and achieve SOTA results. Therefore, we have included several well-established datasets for code summarization, namely TL-CodeSum \cite{hu2018deepcode}, DeepCom \cite{hu2018deep}, FunCom \cite{leclair2019funcom}, and FunCom-50 \cite{chai2022pyramid}. The FunCom-50 dataset was used by PA-Former~\cite{chai2022pyramid}, but with a number of samples filtered out from FunCom, approximately 50\% of the data, to fit their architecture. To maintain a level of fairness, we also use the FunCom-50 as a separate dataset in our study. We followed the original dataset's partition for training, testing, and validation. The statistics for each dataset can be found in the Appendix. 

We pre-process the data by replacing string literals in the code with the generic token \textit{<str>}. The first sentence of the method's description was used as the ground truth summary and all texts were set to lowercase. We trained two Byte-Pair Encoding tokenizers, one for the source sequences and one for the target sequences, for each dataset.


\textbf{Baselines.} In our evaluation, we primarily compare our approach with CAST~\cite{Shi2021CASTEC} and PA-Former~\cite{chai2022pyramid}, as they are the state-of-the-art methods for code summarization. Additionally, we also include other baselines, grouped by code representation or neural architecture, such as \textbf{sequence-based models }(NCS~\cite{ahmad2020ncs} and CodeNN~\cite{iyer2016codenn}), \textbf{structure-based and tree-based models} (Code2seq~\cite{alon2018code2seq}, HybridDRL~\cite{wan2018HybridDrl}, AttGru and ASTAttGru~\cite{leclair2019funcom}, HDeepCom~\cite{hu2020deep}, TPTrans~\cite{peng2021tptrans}, TreeLSTM~\cite{TreeLSTM}, CodeASTNN~\cite{Shi2021CASTEC}, SiT~\cite{hongqiu2021sit}), and \textbf{graph-based models} (GREAT~\cite{Hellendoorn2020great}). 


\textbf{Metrics.} We employ BLEU \cite{Papineni2022bleu}, Meteor \cite{banerjee2005meteor}, Rouge-L \cite{lin2004rouge}, Cider \cite{vedantam2015cider} and F1-score, which are commonly used as evaluation metrics for code summarization.

\textbf{Results.} The results presented in Table~\ref{tab:code-summarization-cast} indicate that {\methodname} outperforms the CAST method by a significant margin on both the TL-CodeSum and FunCom datasets. Specifically, {\methodname} achieved an average improvement of 2.26 BLEU scores over CAST. The comparison with PA-Former on the DeepCom and Funcom-50 datasets, presented in Table \ref{tab:code-summarization-pa}, also shows a substantial margin of improvement for {\methodname}, with an average improvement of 5.175 BLEU scores over PA-Former. This is noteworthy as PA-Former, which is currently considered the state-of-the-art baseline, only outperforms CAST by an average of 1.36 BLEU scores. In conclusion, our evaluations demonstrate that {\methodname}, which utilizes a hierarchical-based architecture and dependencies information, significantly improves performance in code summarization tasks. Additionally, it is worth noting that our evaluations are more extensive than those of PA-Former and CAST, as they only evaluated on two datasets each. This further highlights the absolute superiority of HierarchyNet in comparison to the other methods.

\section{Model Analysis} \label{sec:add_analysis}

We conducted an ablation study to examine the impact of each component of our model design on code summarization performance. We start with the AST-sequence layer and gradually add the Subtree and Graph layers, evaluating the effect of different types of semantic edges. The results in Table~\ref{tab:ablation-bottomup} show the significance of incorporating all three layers in our proposed HCR framework, with a few notable points:

\begin{itemize}[noitemsep]
	\item Using just the AST-sequence layer results in poor performance. Adding the Subtree and Graph layers improves performance incrementally.
	\item Our results using only the AST-edge (row 3) at the Graph level outperform CAST (as seen in Table~\ref{tab:code-summarization-cast}). This is achieved without the need for additional semantic edges, and supports our hypothesis that our hierarchy is superior to the hierarchy proposed by CAST.
	\item The CD and DF edges are equally important as they have a similar impact on performance in rows 4 and 5. However, the NS edges are less vital as their absence has the smallest effect on performance, as seen in row 6 compared to the full model (row 10). Nevertheless, any missing edges will still result in lower performance, as seen in rows 6, 8, and 9 compared to the full model (row 10).
\end{itemize}

\section{Conclusion} \label{sec:conclusion}


This paper proposes a novel framework for code summarization that integrates the use of Heterogeneous Code Representations (HCRs) and HierarchyNet, a neural architecture specifically designed to process HCRs. 
Our HCRs capture essential code features at
lexical, syntactic, and semantic levels by abstracting coarse-grained code elements into a high-level layer and incorporating fine-grained program elements in a lower-level layer.
The HierarchyNet is designed to process each layer of the HCR separately, enabling the accumulation of information from the fine-grained level to be represented as input on the coarse-grained level. The results of our evaluations demonstrate the significant superiority of our approach over other state-of-the-art techniques for code summarization, such as CAST, PA-Former, and others.

\pagebreak

\section*{Limitations}
Our approach presents the opportunities for improvement.

\begin{enumerate}[leftmargin=*]
	\item First, our Heterogeneous Code Representations (HCRs) with coarse-grained semantic edges have proven effective for code summarization. However, there may be potential for further enhancement by exploring alternative options for cross-layer semantic edges, such as connecting nodes at the fine-grained level with nodes at the coarse-grained level. This could be beneficial for other code representation learning tasks, such as variable name prediction \cite{Allamanis2018} and data flow analysis using neural models \cite{gu2021detecting}. Our next step is to conduct further research on extending HCRs to include these alternative options and evaluating their performance on other code representation learning tasks.
	\item Second, while {\methodname} effectively processes the HCRs, there is room for further optimization. We chose the layers in the {\methodname} based on heuristics, resulting in the HGT being the best option for processing graphs. At the subtree-level, we chose the TBCNN as it is more computationally efficient compared to other state-of-the-art methods for processing ASTs, such as TreeCaps~\cite{bui2021treecaps}. However, our approach can be considered a framework rather than a single neural model, so other advancements in tree- or graph-based models or sequence-based models can easily be incorporated to improve performance.
	\item Finally, we did not provide any analysis on the explainability of our model. Explainability is an important aspect of code learning models~\cite{bui2019autofocus,bielik2020adversarial,zhanggenerating,rabin2021generalizability}, and is crucial for the real-world usage of practitioners in code summarization. Our current model design has the potential to support explainability in the future, as the inputs of the high-level layer are computed based on the attention aggregation mechanism, with each input being assigned an attention score. These attention scores can be used to visualize and explain the importance of code elements in a hierarchical way. We will explore this extension as a future work.
\end{enumerate}

\section*{Ethics Statement}
Our framework aims to revolutionize the way software is modeled by taking a new approach with a broader impact in the field. While language models for code have shown impressive performance and have the potential to boost developer productivity, they still face challenges with computational cost and memory consumption. For example, when modeling code and software at the repository level, such as on Github, the AI framework must consider the context of the current code being edited, as well as additional contexts from other files or API calls from external libraries. This is a dependency on a larger scale level in the context of software modeling. Currently, language models typically only model code within the scope of a function or within a single file for tasks such as code summarization or generation. However, this limitation may not be due to the language model itself, but rather the infrastructure of supported IDEs and the software modeling approach. We propose a more realistic way to represent programs as "repository=>file=>class=>function=>statement=>token." The simplest way to model such a hierarchy is to treat them all as a very large sequence and use large language models to model it, but this results in large memory consumption and expensive computational costs. A more affordable approach is to represent large software as modules, where each module can be represented differently at each level. Each layer may require dependency analysis or not, depending on its characteristics. For example, the semantic edges used to connect components clearly differ at each level, requiring careful design of them. Existing approaches to representing the entire program as a graph will fail in this case because the set of semantic edges are designed the same for all nodes without treating them differently. Also, each of the modules can be preprocessed separately on different computing units and aggregated later to achieve efficient computation cost and save memory. We have already demonstrated efficacy when modeling the program at three levels: "function=>statement=>token" and plan to extend this further. 
Our natural way of structuring the source code hierarchically is also aligned well with the advances in programming language and software engineering research in program representations. 
We believe our solution can be viewed as a framework and opens up a new research direction for representing software.


\bibliography{custom}
\bibliographystyle{acl_natbib}

\appendix
\section{Appendix}
\label{sec:appendix}

\subsection{Code Hierarchy}
\label{sec:appendix-code-hierarchy}
\subsubsection{Rule Specification}
As presented in Section \ref{sec:code-hierarchy-rep}, we extract a set of subtrees from an AST by carefully designed rules to form the second level and build the AST-level of our HCRs. Since there is the page limit, we introduce these rules in Table \ref{tab:rule-specs}.
\begin{table*}[thp!]
	\centering
	\fontsize{8.5}{8.5}\selectfont 
\begin{tabularx}{\textwidth}{llX}
		\toprule
		method\_declaration & : & \textit{method\_header} '\{' statement* '\}' \\
		statement & : & simple\_statement | compound\_statement \\
			\textit{simple\_statement}& : & variable\_declaration | return\_statement | method\_invocation | 
			assignment\_statement |                            break\_statement | continue\_statement | throw\_statement | expression\_statement \\
			compound\_statement & : & if\_statement | while\_statement | for\_statement | dowhile\_statement | switchcase\_statement | try\_statement | try\_with\_resources\_statement \\
			if\_statement & : & 'if' '(' expression ')' '{' statement* '}' ('else' 'if'  '(' expression ')' '\{' statement* '\}')?  ('else' '\{' statement* '\}')? \\
			while\_statement & : & 'while' '(' expression ')' '\{' statement* '\}' \\
			for\_statement & : & 'for' '(' variable\_declaration? ';' expression? ';' expression? ')' '\{' statement* '\}' \\
			dowhile\_statement & : & 'do' '\{' statement* '\}' while '(' expression ')' \\
			switchcase\_statement & : & 'switch' '(' expression ')' '\{' ( 'case' expression ':' '\{' statement* '\}')* '\}' \\
			catch\_clause & : & 'catch' '(' identifier identifier ')' '\{' statement* '\}' \\
			finally\_clause & : & 'finally' '\{' statement* '\}' \\
			try\_statement & : & 'try' '\{' statement* '\}' catch\_clause* finally\_clause? \\
			try\_with\_resources\_statement & : & 'try' '(' resource\_specification ')' '\{' statement* '\}' catch\_clause* finally\_clause? \\
			\textit{expression} & : & ternary\_expresison | assignment\_expression | instanceof\_expression | cast\_expression | binary\_expression |
			unary\_expression | resource\_specification | update\_expression | primary\_expression \\
			primary\_expression & : & literal | this | identifier | field\_access | array\_access \\
			\midrule
			INPUT & : & method\_declaration \\
			OUTPUT & : & {method\_header} | simple\_statement | expression \\
		\bottomrule
\end{tabularx}
\caption{The rule specification to split AST into subtrees}
	\label{tab:rule-specs}
\end{table*}
\subsubsection{Dependency Types at the Graph Level}
As presented in the subsection \ref{subsec:semantic-level}, we build a graph with the four different types of dependencies: AST (structural), control-dependence, data-dependence, and next-subtree,  to model the relationships among subtrees. Below are the details of each type.
\begin{enumerate}[leftmargin=*]
	\item \textbf{Structural Dependency (AST-E)} $T^\prime$ consists of a mixture of original AST nodes from $T$ and new subtree nodes, the AST edges serve as the syntactical representation of code. In Figure ~\ref{fig:overview}, AST edges are denoted as green arrows.
	\item \textbf{Control Dependency (CD-E)}: Control dependency occurs when a program instruction executes if evaluating the preceding instruction permits its execution. In Figure \ref{fig:overview}, the subtree \textit{max = b} is \textit{control-dependent} on the subtree \textit{a < b}, therefore, \textit{max = b} connects with \textit{a < b} via a control-dependence edge (red arrows in Figure~\ref{fig:overview}).
	
	\item\textbf{Data Dependency (DF-E)}: Data flow indicates how the values of variables change as a program is executed. 
 For example, there is a data-flow edge (yellow dashed arrow) between the two nodes \textit{int max = a} and \textit{return max}.
	
	\item\textbf{Next-Subtree (NS-E)}: This type  represents the sequential order of subtrees, not the execution ones. For example, the subtree \textit{max = b} is written right after \textit{a < b}, but \textit{max = b} may not be necessarily executed after \textit{a < b}. Thus, the node \textit{a < b} connects with \textit{max = b} through a next-subtree edge (blue dashed arrows in Figure~\ref{fig:overview}).
\end{enumerate}

In summary, the AST edges are utilized to reflect the AST's structure. To encode the order of subtrees in source code, an NS edge is used to connect a subtree with the subsequent one. 
CD- and DF- edges are incorporated into the graph based on static program analysis and serve as inductive bias. It is clear that including CD- and DF-edges helps shorten the distance between pairs of nodes, which lessens the long-term dependency problem.

\subsubsection{Comparing with other methods}

\begin{figure*}[t]
	\centering
	\includegraphics[width=\linewidth]{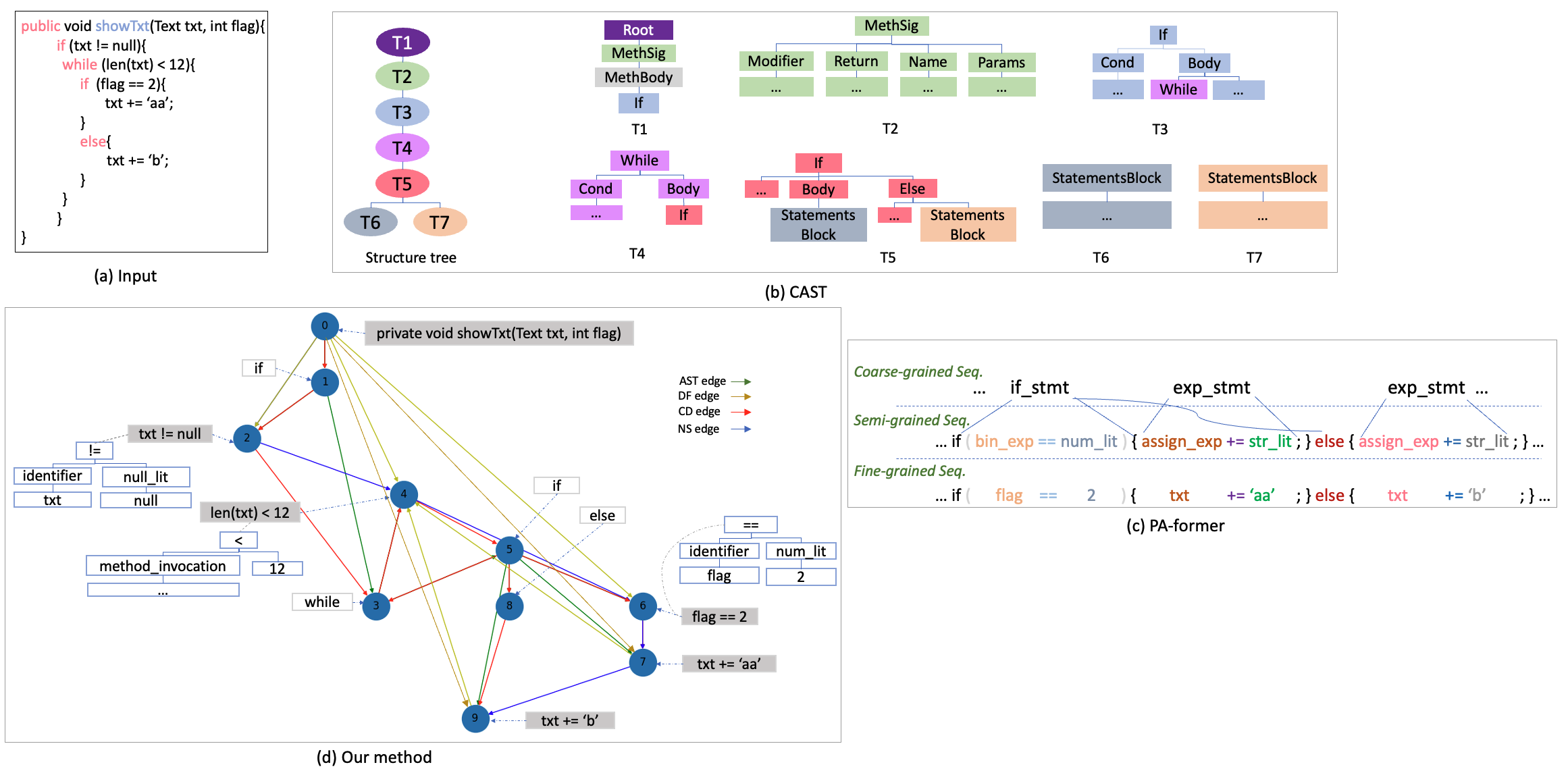}
	\caption{The representations of CAST, PA-former, and our method. For visualization purposes, in our method, we just show the graph-level and some of the subtrees (gray nodes in the graph) as tree structure at the subtree-level. Similarly, we also show a portion of PA-former's representation}
	\label{fig:rep_comparison}
\end{figure*}

PA-former and CAST are the recent state-of-the-art method in code summarization. They also divide an AST into smaller parts to create a hierarchy; however, they are lack of capturing dependencies between code elements. This section aims to explain the departure points of our code hierarchy approach from their models.

CAST's main goal is to re-structure the AST in the top-down fashion. It identifies the composite statements (i.e., \texttt{if}, \texttt{while}, \texttt{for}, \texttt{switch}), and considers them as the subtrees. The consecutive statements are groups in a block and new AST subtree is created for such a block. It recursively processes in the same way for each composite statement one level at the time until there is no more composite statement at the lowest level. It also includes a tree structure that depicts the ancestor-descendant relations among substrees. In comparison, this approach presents two major challenges. First, because a composite structure (a subtree) can take up a large portion of the function and each subtree is encoded in an embedding vector, such a vector is unlikely to compress sufficient information from a large subtree. Second, the structure tree can deteriorate into a degrading tree, which seriously suffers from the long-term dependency problem. Figure~\ref{fig:rep_comparison}a) and b) display an example of source code and the corresponding CAST representation. Because at each level, there exists only one single composite statement, the tree-structure at the highest level deteriorates into a straight line. Unlike that, the code hierarchy in our approach (Figure~\ref{fig:rep_comparison}d) does not have the degrading structure problem because we divide the structure into three separate levels as explained.

 For PA-former, from source code, it constructs {\em three sequences} of sub-tokens, grammar units and logic units, corresponding to the fine-grained, semi-grained and coarse-grained levels. An unit at the higher level is considered to have a corresponding sequence of the units at the immediate-lower level. Moreover, it has a cross-granularity interaction layer by concatenating the three levels and feeds them to a multi-head attention layer. A key drawback is that the three levels are essentially sequences, which makes it harder for a model to learn the structural information and dependencies among program elements (Figure~\ref{fig:rep_comparison}). 
 

In our approach, we use a heterogeneous representation that is suitable to different levels of abstraction. At the lexical level, we have the code sequences that correspond to subtrees. For example, in Figure~\ref{fig:rep_comparison}d), 
the node for line 2 corresponds to the text \texttt{txt != null}. At the syntactic level, the AST structure is used as it naturally represents the syntactic structure. At the semantic level, the dependencies are represented among the statement and expression nodes. An noteworthy point in our approach is that we choose a heterogeneous representation with the natural representations for individual levels, and design a complex model that is suitable to process that representation. In contrast, CAST and PA-former do not create a naturally hierarchical structure of program elements, instead making it suitable to their chosen model.  
In our approach, we design a heterogeneous model that integrates sequence-based transformer, tree-based encoder, and graph-based transformer in which each of them handles the representation at lexical, syntactic, and semantic levels. 

\subsection{Analysis}
\label{sec:appendix-case-study}
\subsubsection{Network Architecture}
\begin{table}[!htp]
	\resizebox{\columnwidth}{!}{
		\begin{tabular}{  l c c c c} 
			\toprule
			Method & BLEU & Meteor & Rouge-L & Cider\\
			\midrule
			\methodnamews & \textbf{48.01} & \textbf{30.30} & \textbf{57.90} & \textbf{4.20} \\
			\textit{w/o Hierarchy-aware} &  46.63& 29.49 & 56.63 & 4.03\\
			\textit{w/o TokenIndexSelector} & 45.70 & 28.39 & 55.06 & 3.93\\
			\bottomrule
	\end{tabular}}
	\caption{{Ablation study of the network architecture on TL-CodeSum}}
	\centering
	\label{tab:ablation-study-network}
\end{table}
Besides a thorough ablation study of the impact of each component in our proposed Heterogeneous Code Representation on the performance (Section \ref{sec:add_analysis}), we also conduct another ablation study to demonstrate the significance of our proposed layers in Hierarchy Net, including Hierarchy-Aware Cross-Attention (abbreviated as Hierarchy-Aware) and TokenIndexSelector, on the TL-CodeSum dataset. The result (Table \ref{tab:ablation-study-network}) shows that the removal of any of these components significantly degrades performance. This confirms that the Transformer architecture alone is not sufficient to encode both textual and structural/semantic meanings of code, thus highlighting the importance of explicitly integrating semantic and structural information using Hierarchy-Aware Attention. Additionally, we found that removing TokenIndexSelector has a negative impact on performance, which is likely due to the redundant information in the sequence provided to the Decoder. 

To show the effectiveness of the serial decoding with the two consecutive cross attention in the Decoder, we compare to two alternatives that just use a cross attention in the Decoder. Specifically, the first option calls for utilizing the Gating layer's output. The other way is concatenating the TokenIndexSelector's output, HGT's output and GraphAggr's output into single large sequences, which are then fed to the Decoder. As shown in Table \ref{tab:ablation-study-decoding}, more information employed in the Decoder leads to the better performance compared to only Gating layer's output. Furthermore, combining our proposed code hierarchy representation with the serial decoding achieves the highest results.
\begin{table}[!htp]
	\resizebox{\columnwidth}{!}{
		\begin{tabular}{  l c c c c} 
			\toprule
			Decoding strategy & BLEU & Meteor & Rouge-L & Cider\\
			\midrule
			serial decoding & \textbf{48.01} & \textbf{30.30} & \textbf{57.90} & \textbf{4.20} \\
			{only Gating layer's output} & 45.34& 28.28& 55.33 & 3.89\\
		{concat} & 47.22& 29.41& 56.45 & 4.10\\
			\bottomrule
	\end{tabular}}
	\caption{{Ablation study of decoding strategy on TL-CodeSum}}
	\centering
	\label{tab:ablation-study-decoding}
\end{table}

\subsubsection{Qualitative Example}
In addition to the quantitative assessments in Section \ref{sec:add_analysis} that demonstrate the necessity of each component in our proposed hierarchical representation of the code, we also provide two examples to visually see the incremental efficacy of each component to code understanding. 

\textbf{Example 1.} Given the above source code (Figure \ref{fig:quanli-ex-2}), the outputs of the different variants of the code representation are shown in Table \ref{tab:quanli-ex-2}. We make following observations:
\begin{figure}[!htp]
	\centering
	\includegraphics[width=\linewidth]{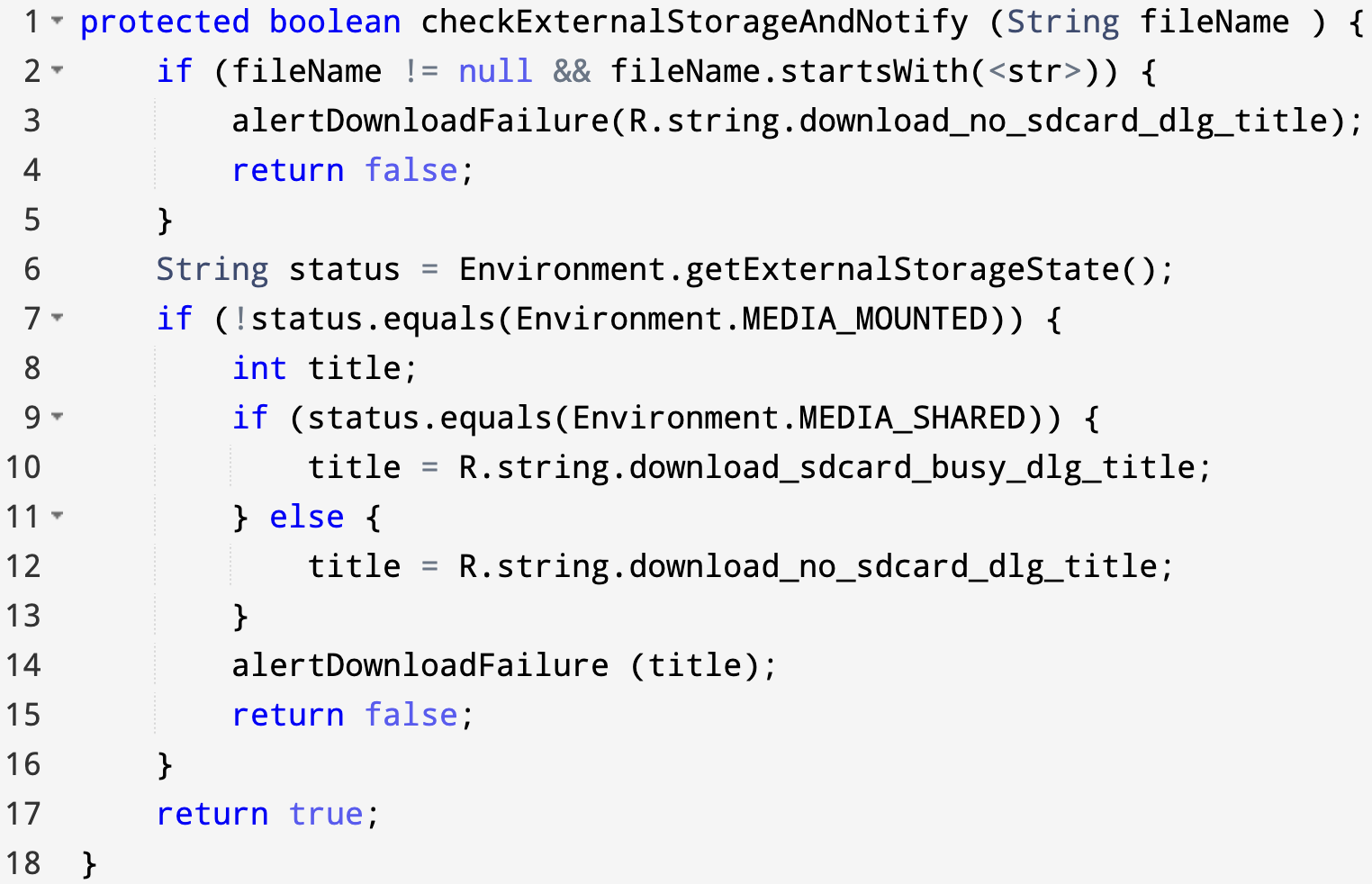}
	\caption{\small Code example 1} 
	\label{fig:quanli-ex-2}
\end{figure}

\begin{enumerate}
	\item The first two experiments show the models misunderstand the code,
	\item When representing code that incorporates the graph-level but with only AST edges, the results are improved in comparison with the formers. Nevertheless, the summary is incomplete due to the lack of complex relationships between subtrees, i.e., data flow and control dependence,
	\item The best results come from the model with a full of layers.  This is due to the fact that the piece of code (Figure \ref{fig:quanli-ex-2}) is rather complex when it has two large \textit{if} blocks, so we cannot overlook control dependencies among subtrees. Besides, there are data dependencies through the variables \textit{status} and \textit{title}. Our proposed representation explicitly depicts these dependencies, making our approach good enough to comprehend the source code.
\end{enumerate}
\begin{table}[!htp]
	\resizebox{\columnwidth}{!}{
\begin{tabularx}{\textwidth}{llX}
			\toprule
			ID & Options & Sentence\\
			\midrule
			1 & Tokens &  check storage percentage , and sends it to the classify storage bar for any other download \\
			2 & Tokens + Subtrees & check if the external storage file is in the android downloadmanager \\
			3 & Tokens + Subtrees + Graph (only AST edges) &  check if the external storage device \\
			\midrule
			4 &		Tokens + Subtrees + Graph (full of edge types) (ours) &  check the external storage and notify user on error\\
			\midrule
			&		Ground-truth & 		 check the external storage and notify user on error \\
			\bottomrule
	\end{tabularx}}
	\caption{{Example 1 of how including different layers affect on the summary}}
	\centering
	\label{tab:quanli-ex-2}
\end{table}

\begin{figure}[!htp]
	\centering
	\includegraphics[width=\linewidth]{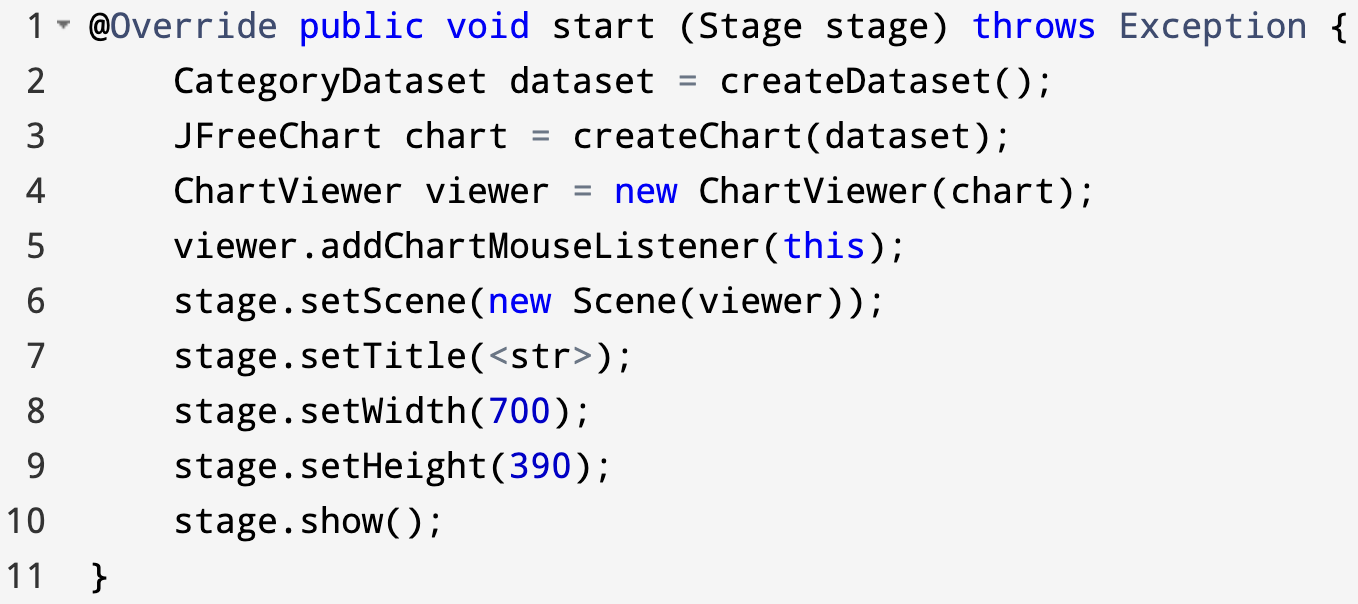}
	\caption{\small Code example 2} 
	\label{fig:quanli-ex}
\end{figure}

\textbf{Example 2.}
Given the above source code (Figure \ref{fig:quanli-ex}), the outputs of the different variants of the code representation are shown in Table \ref{tab:quanli-ex}. We make following observations:
\begin{enumerate}
	\item Only using tokens fails to understand the code,
	\item When subtrees are added, the result shows that the model can capture the actions "create" and "display" of the object "chart viewer",
	\item The result obtained by using only the AST-edge at the Graph-level is better than the previous two experiments. The model produces the word "stage" because the graph models the structural relationships between subtrees. However, the graph misses the order connection between subtrees, so the model fails to produce "displays it",
	\item Our model with a full of options can genererate the summary which matches the ground truth, demonstrating the efficacy of our proposed representation with the three-level hierarchy and the four edge types in the graph.
\end{enumerate}
\begin{table}[!htp]
	\resizebox{\columnwidth}{!}{
		\begin{tabular}{ l l l} 
			\toprule
			ID & Options & Sentence\\
			\midrule
			1 & Tokens &  creates a chart bar chart ( clicked ) \\
			2 & Tokens + Subtrees & creates and displays a chart viewer \\
			3 & Tokens + Subtrees + Graph (only AST edges) &  adds a chart viewer to the stage \\
			\midrule
			4 &		Tokens + Subtrees + Graph (full of edge types) (ours) &  adds a chart viewer to the stage and displays it \\
			\midrule
			&		Ground-truth & 		 adds a chart viewer to the stage and displays it \\
			\bottomrule
	\end{tabular}}
	\caption{{Example 2 of how including different layers affect on the summary}}
	\centering
	\label{tab:quanli-ex}
\end{table}
\subsubsection{Case Study}
To further understand our method, in this section, we conduct a detailed analysis to answer the question, "To what extent does the structural and semantic meaning of code contribute to the performance?".
\begin{figure}[!htp]
	\centering
	\includegraphics[width=\linewidth]{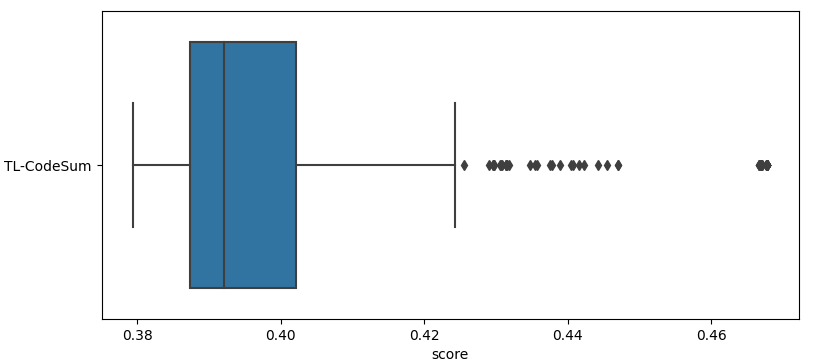}
	\caption{\small The impact of structural and semantic meaning versus textual meaning on the TL-CodeSum} 
	\label{fig:all-dataset-scores}
\end{figure}

In \methodnamews, Gating layer is used to balance the two input sources, textual and structural/semantic information. Thus, we can utilize the scores computed by this layer to show the level of contribution of structure and semantics of code to the result of each sample. For TL-CodeSum, $\lambda$ (Eq. \ref{eq:gating}) is a scalar, so we can directly use this value as the score for each sample. The detailed results are shown in Figure \ref{fig:all-dataset-scores}, which shows that the contributions of structural and semantic meaning are different on the TL-CodeSum. The importance scores of structural and semantic meaning are at least 0.37 while the rest rates belong to the lexical information. As a result, it is crucial to explicitly incorporate these information into the code representation.

\subsection{Experiments}
\label{sec:appendix-experiments}
\subsubsection{Datasets}
\begin{table}[!htp]
	\centering
 \fontsize{8.5}{8.5}\selectfont 
	\begin{tabular}{ l c c c c} 
		\toprule
		Dataset & \#train & \#test & \#summary\\
		\midrule
		TL-CodeSum & 65,845  & 8,199 & 50 \\
		DeepCom & 295,967  & 12,226 & 30\\
		FunCom-50 & 1,014,893 & 53,751 &40\\
		FunCom & 1,911,669 & 104,981 & 20\\
		
		\bottomrule
	\end{tabular}
	\caption{Dataset statistics. \#train and \#test are the number of the training and testing samples, respectively. \#summary is the max sequence length of the summary which is tokenized.}
	\label{tab:statistics-datasets}
\end{table}
In Section \ref{sec:eval} of the main text, we present detailed experiments on the four datasets, TL-CodeSum, DeepCom, FunCom-50, and FunCom, to demonstrate the superiority of our proposed method over other baselines. Table \ref{tab:statistics-datasets} displays the statistics for these datasets. In addition, the validation portions of TL-CodeSum and FunCom have 8231 and 105007 samples, respectively, while the other datasets do not.

\begin{table*}[ht]
	\centering
	\begin{tabular}{lllllll}
		\toprule
		Dataset  & Batch size & Warm-up steps & Max lr & Max epochs & Source/Target Token Vocab Size  \\
		\midrule
		TL-CodeSum & 16 & 30000 & 1e-4 & 70 & 30000 / 18535\\
		DeepCom & 384 & 6400 & 1.2e-4 & 50& 30000 / 30000\\
		FunCom-50 & 256 & 28000 & 1.5e-4 & 60& 32000 / 30000\\
		FunCom & 352 & 32400 & 1.2e-4 & 60& 30000 / 30000\\
		\bottomrule
	\end{tabular}
	\captionof{table}{Hyper-parameters for each dataset.}
	\label{tab:settings-each-dataset}
\end{table*}
\subsubsection{Settings}
All the models have the dimension size of 768, 12 layers for Transformer Encoder and Transformer Decoder, 1 layer for Tree-based CNN, 2 layers for HGT, and 12 heads for all attention layers. Detailed hyperparameters for each dataset are shown in Table \ref{tab:settings-each-dataset}. The TL-CodeSum's experiments are conducted on a server of 4 GPUs of RTX 3090, and the other experiments are run on a server of 2 GPUs of A100. It takes about 1.5h, 0.83h, 3.16h, and 4.8h per epoch for TL-CodeSum, DeepCom, FunCom-50 and FunCom, respectively.

\subsection{Neural Networks}
\label{sec:appendix-neural-networks}
\subsubsection{Tree-based Convolutional Neural Network (TBCNN)}
TBCNN~\cite{mou2016convolutional} is designed to process tree-structure through the tree-based convolution operator. In a TBCNN, there is at least one tree-based convolutional layer. Each layer is a feature detector and has a fixed-depth convolutional window called the kernel, sliding over the entire tree to extract features. Formally, this procedure can be summarized as:
$y = f \left( \sum_{1}^{n} W_{\text{conv}, i} \cdot x_i + b_{\text{conv}} \right)$, where $f$ is an activation function, $W_{\text{conv}, i}$ are the weight matrices, $x_i$ are the vectors of nodes inside the sliding window, and $b_{\text{conv}}$ is the bias. 

 Because of the fixed number of weight matrices, TBCNNs always see all trees as continuous binary trees. This leads $W_{\text{conv}, i}$ to be different for each node. In summary, at each convolutional step, the feature of node $i$ is accumulated by its direct children in a sliding window simultaneously. At the end of this step, the fix-sized embedding of a subtree is computed by using a max pooling operator over all of the nodes in such subtree.
\subsubsection{Heterogeneous Graph Transformer (HGT)}
A HGT layer can be decomposed into three components: heterogeneous mutual attention, heterogeneous message passing and target-specific aggregation. The overall process can be written as: 
\begin{equation}
	\small 
	H^l[t] \leftarrow \underset{\forall s \in N(t), \forall e \in E(s, t)}{\text{Aggregate}}\left( \text{Attention}(s,e,t) \cdot \text{Message}(s,e,t) \right)
\end{equation}
where $N(t)$ is the set of source nodes of node $t$ and $E(s,t)$ denotes all the edges from node $s$ to node $t$.

\textbf{Heterogeneous Mutual Attention} Heterogeneous mutual attention is based on multi head attention in Transformer. For each edge $e = (s, t)$,

\begin{equation}
	\small
	\begin{aligned}  
		\textbf{Attention}(s,e,t) &= softmax \left( \underset{i \in [1,h]}{\Vert}ATT \text{-} head^i \left(s, e, t \right) \right)\\
		ATT \text{-} head^i \left(s, e, t \right)  &= \left( K^i(s) W^{ATT}_{\phi(e)} Q^i(t)^T \right) \cdot \frac{\mu_{\left \langle \phi(e) \right \rangle}}{\sqrt{d}}\\
		K^i(s) &= \text{K-Linear}^i_{\tau(s)} \left( H^{l-1}[s] \right)\\
		Q^i(t) &= \text{Q-Linear}^i_{\tau(t)} \left( H^{l-1}[t] \right)
	\end{aligned}
\end{equation}
The Keys and Queries are calculated based on source node $s$ and target node $t$, respectively. Note that we use a different projector for each node type. Besides, there may exist many relations between a node pair, so a edge-based matrix $W^{ATT}_{\phi(e)}$ for each edge type is added to distinguish disparate semantic relations between such node pairs.

\textbf{Heterogeneous Message Passing} Next, we compute the message for a node pair $e = (s, t)$
\begin{equation}
	\small
	\begin{aligned}  
		\textbf{Message}(s,e,t) &=  \underset{i \in [1,h]}{\Vert}msg \text{-} head^i \left(s, e, t \right) \\
		msg \text{-} head^i \left(s, e, t \right)  &= \text{M-Linear}^i_{\tau(s)} \left( H^{l-1}[s]\right)  W^{msg}_{\phi(e)} \\
	\end{aligned}
\end{equation}
where $\text{M-Linear}^i_{\tau(s)}$ is a matrix used to project the $\tau(s)$-type source node $s$ into $i$-th message vector, and $W^{msg}_{\phi(e)}$ is a edge-based matrix for the edge type $\phi(e)$ to incorporate the edge dependency between node $s$ and node $t$.

\textbf{Target-specific Aggeregation} Eventually, we aggregate the messages with the corresponding attention scores, then the target node $t$ is updated based on new information from all of its neighbors and its previous state.
\begin{subequations}
	\small
	\begin{align}  
		\tilde{H}^l[t] &= \underset{\forall s \in N(t)}{\sum}\left( \textbf{Attention}(s,e,t) \cdot \textbf{Message}(s,e,t) \right)\\
		{H}^l[t] &= f \left(\text{A-Linear}^{\tau(t)}\left( \phi \left( \tilde{H}^l[t] \right) \right), H^{l - 1}[t] \right) \label{eq:agg-node}
	\end{align}
\end{subequations}
where $f$ is a Feed Forward layer.

\end{document}